**Application of the fluctuation theorem to motor proteins: from $F_1$-ATPase to axonal cargo transport by kinesin and dynein**

Kumiko Hayashi

Department of Applied Physics, Graduate School of Engineering, Tohoku University, Sendai, Japan

Correspondence: kumiko@camp.apph.tohoku.ac.jp

**Abstract**

The fluctuation theorem is a representative theorem in non-equilibrium statistical physics actively studied in the 1990's. Relating to entropy production in non-equilibrium states, the theorem has been used to estimate the driving power of motor proteins from fluctuation in their motion. In this review, usage of the fluctuation theorem in experiments on motor proteins is illustrated for biologists, especially those who study mechanobiology, in which force measurement is a central issue. We first introduce the application of the fluctuation theorem in measuring the rotary torque of the rotary motor protein $F_1$-ATPase. Next, as an extension of this application, a recent trial estimating the force generated during cargo transport *in vivo* by the microtubule motors kinesin and dynein is introduced. Elucidation of the physical mechanism of such transport is important, especially for neurons, in which deficits in cargo transport are deeply related to neuronal diseases. Finally, perspectives on the fluctuation theorem as a new technique in the field of neuroscience are discussed.

**Introduction**

The fluctuation theorem is a representative theorem in non-equilibrium statistical physics that quantifies the probability of entropy production in non-equilibrium systems (Ciliberto et al. 2010; Evans et al. 1993). As fluctuating behavior is a typical property of bio-molecular motion, the fluctuation theorem has been applied to single-molecule experiments on DNA/RNA hairpins to estimate free energy differences between folded and unfolded states (Alemany et al. 2012; Camunas-Soler et al. 2017; Collin et al. 2005). In the 2000's, observations using optical microscopes and optical tweezers in nano- and micro-scale systems matched the applications of the fluctuation theorem.

In the next decade, application of the fluctuation theorem progressed, because many variations of the theorem can be derived from the original equation (Ciliberto et al. 2010). One such variation has been used to measure the driving power of motor proteins (Hayashi et al. 2010). Through the theorem, the fluctuating motion of a molecular motor system, which is subject to thermal noise, is related to heat production by the driving power of motors. Thus far, the theorem has been applied to single-molecule experiments on rotary motor proteins, such as $F_1$-ATPase and $V_1$-ATPase (Enoki et al. 2015; Kim et al. 2011; Kishikawa et al. 2014; McMillan et al. 2016; Ueno et al. 2014; Usukura et al. 2012; Watanabe et al. 2014). From the statistical property of fluctuation, the torque of these rotary motors was non-invasively estimated without any external stall torque.

Because non-invasive measurement is desirable for measurements of living cells, which are sensitive to external stimuli, studies have begun to apply the theorem to estimate the force of motor proteins *in vivo*. Non-invasive force measurements are often used in mechanobiology, an emerging field of science; for example, the force on the outer surface of cells or on the plasma membrane can be measured by traction force microscopy (Polacheck and Chen 2016). Fluorescent protein-based biosensors for measuring force or tension at the cellular level have been developed using Förster resonance energy transfer (Guo et al. 2014; Meng and Sachs 2012; Meng et al. 2008). In this review, using the fluctuation theorem, we introduce the challenge of measuring the force generated during cargo transport in living neurons by microtubule motor proteins, such as kinesin and dynein. Because deficits in intracellular transport, particularly in neurons, are related to neuronal diseases, such as Alzheimer's, Parkinson's, and Huntington's disease (Chiba et al. 2014; Encalada and Goldstein 2014), elucidation of the physical mechanism of *in vivo* cargo transport is significant.

In the following paragraphs, we first review the application of the fluctuation theorem to $F_1$-ATPase. We then explain the modifications to the theorem necessary for studying *in vivo* cargo transport. Finally, application of the theorem to synaptic vesicle precursor transport in the axons of *Caenorhabditis elegans* worms is introduced as a concrete example of the modified fluctuation theorem. As a perspective, further applications in the neuroscience field are discussed.

**Application of the fluctuation theorem to $F_1$-ATPase**

**$F_1$-ATPase**

$F_1$-ATPase ($F_1$) is a rotary motor protein and a part of $F_oF_1$-ATPase/synthase (Noji et al. 1997; Okuno et al. 2011). The minimum complex required for motor function is the $\alpha_3\beta_3\gamma$ subcomplex, in which the γ subunit (rotor) rotates in the $\alpha_3\beta_3$ ring upon ATP hydrolysis (Fig. 1a). The three catalytic β subunits hydrolyze ATP sequentially and cooperatively. The push-pull motion of the β subunits produces torque, rotating the γ subunit.

**Single-molecule assay of $F_1$**

In a single-molecule assay (Hayashi et al. 2010), the rotation of the γ subunit was observed by optical microscopy as the rotation of a streptavidin-coated bead attached to the biotinylated γ subunit (Fig. 1b). The histidine-tagged $F_1$ was tightly attached to a $Ni^{2+}$-NTA-coated glass surface. The rotational angle (θ) was calculated from recorded images of the bead. In Fig. 2a, θ is plotted for the case of 1 mM ATP and wild-type $F_1$. The ATP binding dwell is very short at high [ATP] (< 0.1 ms at 1 mM), and the time constants of dwells for ATP hydrolysis and for product release are also short (approximately 1 ms for wild-type $F_1$). The rotation appears continuous and the angular velocity is almost constant for several seconds. Fluctuating behavior of θ was observed at a high recording rate (2000 frames per second (fps)) (Fig. 2b, top).

**Fluctuation theorem**

In order to measure the rotary torque (*N*) of $F_1$, we apply the fluctuation theorem, expressed as

$$N = \frac{\ln[P(\Delta\theta)/P(-\Delta\theta)]}{\Delta\theta} \cdot k_{\mathrm{B}}T \qquad (1)$$

where $\Delta\theta = \theta(t+\Delta t) - \theta(t)$, $P(\Delta\theta)$ is the probability distribution of $\Delta\theta$, $T$ is the temperature of the environment ($T$=25 ℃), and $k_{\mathrm{B}}$ is the Boltzmann constant (See Ref. (Hayashi et al. 2012) for the derivation of Eq. (1).) Noting that the fluctuation theorem was originally the theorem for entropy production, $N\Delta\theta$ and $N\Delta\theta/T$ are heat and the entropy production of the $F_1$ system generated during $\Delta t$, respectively. We also note that constant torque is assumed by Eq. (1); that is, the torque is dependent on $\theta$ (Toyabe et al. 2012). However, this assumption was reasonable based on the precision required in the comparison experiments between wild-type and mutant $F_1$, which are introduced in the later section "*Further applications*."

For the continuous rotation of $F_1$, $P(\Delta\theta)$ and $\ln[P(\Delta\theta)/P(-\Delta\theta)]$ are plotted for the cases $\Delta t$ = 2.5–10 ms. The constant torque $N$ calculated from Eq. (1), which is the slope of the graphs in Fig. 2b (bottom), is plotted as a function of $\Delta t$ in Fig. 2c. After some relaxation time of the system, $N$ is convergent to the constant value $N^*$.

Previously, the torque was estimated by using fluid mechanics, as follows:
$$N_\Gamma = \Gamma\omega \qquad (2)$$
where $\omega$ is the angular velocity and $\Gamma$ is the friction coefficient of the probe bead. Here, $\Gamma$ is calculated to be $40\pi\eta a^3$ ($\eta$: viscosity of the medium, $a$: radius of the bead) for the bead duplex according to fluid mechanics equations (Hayashi et al. 2010; Howard 2001). For the $F_1$ system, it was found that $N^*$ estimated from Eq. (1) corresponded to $N_\Gamma$ (Hayashi et al. 2010). For example, when the probe is a duplex of 470 nm polystyrene beads, $N^* = 31 \pm 2.6$ (SEM) pN nm and $N_\Gamma = 31 \pm 3.3$ pN nm. That $N^*$ and $N_\Gamma$ agreed within the precision of the experiment supports the validity of Eq. (1). (Note that the torque was estimated to be about 40 pN nm previously in Ref. (Noji et al. 2001). The value of $N$ estimated using Eq. (1) ranged about from 30 pN nm to 40 pN nm depending on the experimental conditions such as sizes of probes and ATP concentration (Hayashi et al. 2010).)

Information on the shape of a probe and its attachment to $F_1$ is needed when the friction coefficient ($\Gamma$) is calculated using Eq. (2). However, it is often difficult to measure such information precisely. As an example of the difficulty in calculating $\Gamma$, images of irregularly shaped magnetic beads are shown in Fig. 3a. The magnetic beads were often fractured to make them smaller. Additionally, it was difficult to measure how the nano-rods recently used as probes in Ref. (Enoki et al. 2015) attached to $F_1$ (Fig. 3b), and the value of $\Gamma$ largely depends on this attachment.

**Further applications to rotary motors**

Because Eq. (1) does not require probe information (shape, size, or other parameters) to calculate $\Gamma$ (unlike Eq. (2), which does require these things to estimate torque), Eq. (1) is commonly used to study rotary motors. Thus far, it has been used to study many kinds of rotary motor proteins, such as mutants of $F_1$ and $V_1$-ATPase. Torque values of wild-type and mutant *Thermus thermophilus* $V_1$ have been compared to investigate the function of the F-subunit of $V_1$ (Kishikawa et al. 2014). The effect of the DELSDED loop of $F_1$ on torque generation has been investigated (Tanigawara et al. 2012; Usukura et al. 2012). Torque values in cyanobacterial $F_1$ have been measured to elucidate the redox regulation mechanism of chloroplast $F_1$ (Kim et al. 2011). Torque values of alanine (near the ATP-binding site)-substituted mutants of *Thermophilic Bacillus* PS3 $F_1$ have been investigated (Watanabe et al. 2014). Torque generation by *Enterococcus hirae* $V_1$ has been investigated (Ueno et al. 2014). Torque measurements have been successfully made using nano-rods as probes (Enoki et al. 2015). The torque generation of *Caldalkalibacillus thermarum* TA2.A1 $F_1$-ATPase has been investigated (McMillan et al. 2016).

Bacterial flagella are also rotary motors whose torque is estimated by using Eq. (2) (Beeby et al. 2016; Inoue 2017; Sowa et al. 2005). It is future issue to apply Eq. (1) to such high torque motors.

**Application of the fluctuation theorem to axonal cargo transport**

**Cargo transport by kinesin and dynein**

Microtubules extend throughout eukaryotic cells, and cargo, such as mitochondria and endosomes, which contain many kinds of molecules needed for cellular activity, are transported by ATP-dependent motor proteins, such as kinesin and dynein, along microtubules (see Fig. 4a for the case of a neuron). In cells, the force (*F*) acting on cargo during transportation is not obtained from the velocity ($v$) of the cargo from the equation $F=\Gamma v$, because the friction coefficient ($\Gamma$) of the cargo is difficult to estimate due to the diffraction limit of microscopes and the unknown viscosity of the cytosol. Force measurement using the fluctuation theorem, which does not require $\Gamma$, is useful in this situation.

**Violation of the fluctuation dissipation theorem**

In the case of $F_1$, the fact that the torque ($N$) estimated by the fluctuation theorem (Eq. (1)) corresponds experimentally to the torque ($N_\Gamma$) measured using the friction coefficient (Eq. (2)) (Hayashi et al. 2010) indicates that the fluctuation dissipation theorem was valid for the continuous rotation of $F_1$, because the following fluctuation dissipation theorem was derived from Eqs. (1) and (2) under the condition $N = N_\Gamma$

$$D = \frac{k_\mathrm{B} T}{\Gamma} \qquad (3)$$

Here we used the functional form of $P(\Delta X)$ derived from the diffusion equation

$$P(\Delta\theta) = \exp\left(-\frac{(\Delta\theta - \omega\Delta t)^2}{4D\Delta t}\right) / \sqrt{4\pi D \Delta t} \qquad (4)$$

where $D$ is the diffusion coefficient of a probe bead. Eq. (3) is called the Einstein relation and is one of the fluctuation dissipation theorems, which are valid in near equilibrium.

On the other hand, the fluctuation dissipation theorem (Eq. (3)) was experimentally found to be violated for mitochondria transport in neuron-like cells (PC12 cells) (Hayashi et al. 2013). This implies that the states of mitochondrial transport in PC12 cells were far from equilibrium. Here mitochondria are among the significant organelles transported by microtubule motors in cells. Ref. (Hayashi 2013) compared the diffusion coefficient ($D$) of a mitochondrion calculated from its position with its friction coefficient ($\Gamma$) measured using fluorescence correlation spectroscopy. In the presence of driving force, the complex interaction potentials in the system produced effective noise in addition to the thermal noise acting on the cargo (Fig. 4b). Note that see Ref. (Hayashi and Sasa 2005) for the statistical rule how complex interaction potentials in non-equilibrium are changed to effective noise, effective force and a friction coefficient.

Since thermal noise is dependent on temperature, effective temperature was introduced as a violation factor of the fluctuation dissipation theorem when the theorem was applied to intracellular cargo transport (Hayashi et al. 2018; Hayashi et al. 2017). Setting $X$ as the center position of cargo hauled by microtubule motors, $\Delta X = X(t + \Delta t) - X(t)$, $P(\Delta X)$ as the probability distribution of $\Delta X$, and $F$ as the force acting on the cargo, the temperature of the environment ($T$) is replaced with the effective temperature ($T_{\text{eff}}$) in Eq. (1) as

$$F = \frac{\ln[P(\Delta X)/P(-\Delta X)]}{\Delta X} \cdot k_\text{B} T_{\text{eff}} \qquad (5)$$

For later convenience, we introduce the fluctuation unit $\chi$ as

$$\chi = \frac{\ln[P(\Delta X)/P(-\Delta X)]}{\Delta X} \qquad (6)$$

Note that the concept of effective temperature in non-equilibrium states has been studied in the field of non-equilibrium statistical mechanics (Crisanti and Ritort 2003; Cugliandolo 2011; Hayashi and Sasa 2004) and that effective temperature was also investigated in a single-molecule experiment on DNA hairpins recently (Dieterich et al. 2015). Although the exact physical meaning of effective temperature is still controversial, it is often observed that $T_{\text{eff}} > T$ in non-equilibrium systems.

**Synaptic vesicle precursor (SVP) transport in neurons**
Neuronal morphology necessitates particularly fast cargo vesicle transport for efficient communication between the cell body and distal processes (Hirokawa et al. 2010; Holzbaur 2004). Kinesin superfamily proteins and cytoplasmic dynein haul cargo vesicles anterogradely toward the terminal and retrogradely toward the cell body (Fig. 4a)(Fu and Holzbaur 2014; Hirokawa et al. 2009; Vale 2003). Neuronal communication largely depends on the delivery and receipt of synaptic vesicles in particular. The components of the synaptic vesicles are synthesized in the cell body, and then packaged in synaptic vesicle precursors (SVPs). The SVPs are transported to the synapses, which are located near the axon terminal, by the kinesin superfamily protein UNC-104/KIF1A, which belongs to the kinesin-3 family (Hall and Hedgecock 1991; Okada et al. 1995; Otsuka et al. 1991). Note that UNC-104/KIF1A was originally identified in *C. elegans* (Hall and Hedgecock 1991; Otsuka et al. 1991).

As a concrete example, we introduce the modified fluctuation theorem (Eq. (5)) as applied to SVP transport in neurons of *C. elegans* (Hayashi et al. 2018). Different velocities were observed for the motors *in vivo* in living organisms (several μm/s) (Hayashi et al. 2018) and *in vitro* in single-molecule kinesin experiments (about 1 μm/s) (Schnitzer et al. 2000). Therefore, the motors must have different transport mechanisms under *in vitro* and *in vivo* conditions.

**Fluorescence observation of SVPs in neurons**

Axonal transport of SVPs in DA9 neurons of wild-type *C. elegans* was observed by fluorescence microscopy (Fig. 5). The worms were anesthetized and fixed between cover glasses with highly viscous media to minimize fluctuation from their body movements (see Method in Ref. (Hayashi et al. 2018)). SVPs in the neurons were labelled with green fluorescent proteins (GFPs) (see Method in Ref. (Hayashi et al. 2018)). There were time intervals during which a single SVP could be tracked (red arrows) even though the axons were crowded with many SVPs (Fig. 5b). From the recorded images, the center position ($X$) of a vesicle along an axon was obtained as a function of time ($t$) for anterograde transport (Fig. 6a). Note that anterograde transport was focused on in the investigation of UNC-104 kinesin.

**Calculation of $\chi$ in Eq. (6)**

The analytic target was the constant velocity segment of a transported SVP, as in the case of continuous rotation of $F_1$ (Fig. 2a). When the recording rate was high enough (100 fps (frames per second)), fluctuating behavior of the vesicle was observed even while it exhibited directional motion (inset of Fig. 6a). The origin of the fluctuation in the transported SVPs was thought to be mainly thermal noise, stochastic motion of the motors caused by ATP hydrolysis, and collision of the SVP with other vesicles and the cytoskeleton.

Fig. 6b shows the probability distribution ($P(\Delta X)$) of $\Delta X$ for the case $\Delta t = 100$ ms. It was fitted well by a Gaussian function $P(\Delta X) = \exp(-(\Delta X - b)^2/2a)/(2\pi a)^{0.5}$ in which the fitting parameters $a$ and $b$ correspond to the variance and the mean of the distribution. Using these parameters, $\chi$ in Eq. (6) was calculated as $2b/a$. Note that we used $\chi = 2b/a$ because the ratio $P(\Delta X)/P(-\Delta X)$ in Eq. (6) could not be calculated directly from $P(\Delta X)$ when the portion of $P(\Delta X)$ when $\Delta X < 0$ was not observed, unlike in the single-molecule experiment on $F_1$ (Fig. 2b), in which the recorded rate was much faster (2000 fps)) than the rate of fluorescence observation (100 fps), and $P(\Delta \theta) \neq 0$ when $\Delta \theta < 0$.

Similarly to the case of $F_1$ (Fig. 2c), $\chi$ reaches a constant value $\chi^*$ for a large $\Delta t$ after a certain amount of relaxation time (Fig. 6b). For anterograde transport of SVPs, $\chi$ was calculated for 40 vesicles from 33 individual wild-type worms (Fig. 6b). After applying a clustering analysis (see Ref. (Hayashi et al. 2018) for affinity propagation), the quantal behavior of $\chi$ was observed.

**SVP transport by multiple kinesin motors**

The quantal behavior of $\chi$ was also observed in endosome transport in dorsal root ganglion (DRG) neurons of mice (Hayashi et al. 2017) as well as in SVP transport in DA9 neurons of *C. elegans* (Hayashi et al. 2018). Continuous studies are needed to determine whether the quantal behavior of $\chi$ is universal for *in vivo* cargo transport. The reason that $\chi$ shows quantal behavior can be explained as follows.

From the comparison of $\Gamma v$ ($=F$) and $\chi$ for endosome transport in the axons, Ref. (Hayashi et al. 2017) suggested that $T_{\text{eff}}$ in Eq. (5) was almost constant among different endosomes. Here $\Gamma$ and $v$ are the friction coefficient and velocity of a cargo, respectively, and $F$ is a force generated by motors, which is equal to the drag force, $\Gamma v$ (schematics in Fig. 4b). This implies, from Eqs. (5) and (6),

$$F \propto \chi^* \qquad (7)$$

When there is a proportional relation (Eq. (7)), the quantal behavior of $\chi$ indicates the existence of force-producing units (FPUs). Because force is generated by motors, existence of multiple FPUs implies that multiple motors cooperatively transport a single cargo element (schematics in Fig. 4a).

Previously, by imposing a load on cargo in living cells using optical tweezers, it was found that the distribution of the stall force acting on cargo is quantal, reflecting the number of motors hauling the cargo (Hendricks et al. 2012; Leidel et al. 2012; Mas et al. 2014; Shubeita et al. 2008). The stall force is the maximum force that a motor can exert against a load (see schematics of the force-velocity relation of motors in Fig. 7). Note that large, round cargo, such as the lipid droplets investigated in Ref. (Leidel et al. 2012; Shubeita et al. 2008), can be loaded even in cells by optical tweezers. Because the viscous effect in cells seems much higher than that in water, the drag force acting on cargo ($F$) takes on a value similar to that of the stall force (Fig. 7). Thus, the distribution of $\chi$ as a force indicator was considered to show several clusters as well as the stall force distributions.

**Physical quantities other than $\chi$**

In addition to $\chi$, the velocity and fluorescence intensity of cargo can be non-invasively measured. However, the velocity of cargo is influenced more by its size than by the number of motors carrying it, because cargo sizes are distributed (fluorescence micrographs in Fig. 8a). As a result, the number of motors cannot be estimated from velocity distributions. Indeed, the velocity distributions of the endosomes did not show multiple peaks (see Supplementary Material of Ref. (Hayashi et al. 2017)).

One may think that the number of motors hauling a cargo can be estimated from fluorescent intensities when fluorescence probes are attached to motors directly (Fig. 8b). However, the number of active motors cannot be counted this way, because motors that do not haul a cargo also emit fluorescence. In addition, intrinsic fluorescence exists in cells. Thus, $\chi$ seems a good quantity for non-invasively counting the number of motors cooperatively transporting cargo in neurons.

**Number of motors as a significant indicator**

In a previous study (Niwa et al. 2016), it was reported that the absence of *arl-8* causes decrease in synapse density and mislocalization of synaptic vesicles in the DA9 neurons of *C. elegans*, implying that the lack of ARL-8 reduced anterograde transport capability of SVPs. ARL-8, a molecule related to UNC-104, is an SVP-bound arf-like small guanosine triphosphatase (GTPase) that activates UNC-104 by releasing its autoinhibition (Klassen et al. 2010; Niwa et al. 2016; Wu et al. 2013). However, the physical parameters in vesicle transport that are affected by the absence of ARL-8 remained elusive. In another study (Hayashi et al. 2018), one of these physical parameters, i.e., the number of motors carrying an SVP, was elucidated from the measurement of $\chi$. In the case of an *arl-8*-deleted mutant, the number of motors was decreased (Fig. 9a). Multiple motor transport in axons seems essential to enhance transport capability.

In another study (Chiba et al. 2014), the velocity decrease caused by the deletion of an adaptor protein thought to connect motors to cargo was observed in amyloid precursor protein (APP) transport (Fig. 9b). As amyloid beta is involved in the pathogenesis of Alzheimer's disease, clarification of the number of motors involved in this transport and the physical reason for the velocity decrease is warranted. We believe that investigation into the relations among force, velocity, and motor number using $\chi$ will be key to elucidating the physical mechanisms related to the axonal cargo transport deficit present in neuronal diseases such as Alzheimer's, Parkinson's, and Huntington's disease.

**Physical meaning of effective temperature**

From the force calibration (Hayashi et al. 2017), $k_B T_{eff}$ in Eq. (5) was estimated to be $14 k_B T$ for endosome transport in the axons of mouse neurons and $20 k_B T$ for synaptic vesicle transport in the axons of *C. elegans* (Hayashi et al. 2018). This does not literally mean that the temperature of the cytoplasm is $14 k_B T$ or $20 k_B T$. Rather, the free energy obtained from the hydrolysis of a single ATP is about 20 $k_B T$, which gives scale to the energy of active processes in living cells. Understanding the physical meaning of $T_{eff}$ in axonal transport will require future investigations using theoretical models based on statistical mechanics.

The value of force generated by 1 FPU for anterograde transport was determined to be approximately 5 pN using the values of $k_B T_{eff}$. It has been suggested that 1 FPU corresponds to a dimer of kinesin (Hayashi et al. 2018; Hayashi et al. 2017) because 5 pN was similar to the stall force values obtained through *in vitro* single-molecule experiments (Schnitzer et al. 2000; Tomishige et al. 2002).

**Perspective**

In this review, we discussed recent developments in the experimental application of the fluctuation theorem on motor proteins. Eq. (1) is commonly used to measure the rotary torque of $F_1$-ATPase. Eq. (5) has just started being applied to axonal cargo transport. Although the universality of the quantal behavior of $\chi$, Eq. (7), and the force calibration method suggested in Ref. (Hayashi et al. 2017) are still controversial, the fluctuation unit $\chi$ (Eq. (6)) has potential as a force indicator. Physical measurements of parameters such as force have seldom been performed in living neurons. If the number of motors involved in axonal cargo transport could be non-invasively measured through $\chi$ based on the fluctuation theorem, the basic physical mechanisms of neuronal diseases, neuronal injury, and regeneration might be further understood. We believe that the fluctuation theorem, first studied theoretically in non-equilibrium statistical mechanics, has been expanded as a cutting-edge technique and will be invaluable as a molecular motor counter in future neuroscience research. We hope that non-invasive force measurement using fluctuation such as the one introduced in this paper will become common in the field of mechanobiology.

**Acknowledgement**

We thank the participants of the 2017 annual meeting of the Australian Society of Biophysics for comments on the study. This work was supported by grants (grant numbers JP17gm5810009, and JP17ja0110011) from the Japan Agency for Medical Research and Development (AMED), and a Grants-in-Aid for Scientific Research (KAKENHI) (grant numbers 26104501, 26115702, 26310204, and 16H00819) from the Ministry of Education, Culture, Sports, Science, and Technology.

**Compliance with ethical standards**

**Conflicts of interest**

Kumiko Hayashi declares that she has no conflicts of interest.

**Ethical approval**

This article does not contain any studies with human participants or animals performed by any of the authors.

# Figure 1

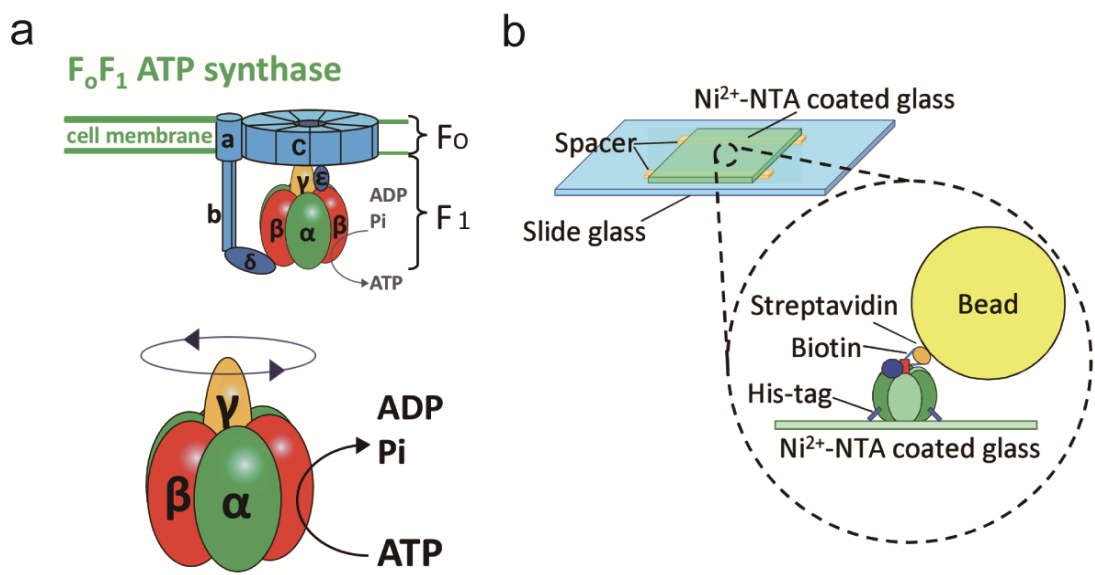

**Fig. 1** Introduction to $F_1$-ATPase. **a** $F_1$-ATPase is a part of $F_oF_1$-ATP synthase. The γ subunit (rotor) rotates in the $α_3β_3$ ring upon ATP hydrolysis. **b** In a single-molecule experiment on $F_1$ (Hayashi et al. 2010), histidine-tagged $F_1$ was attached to a $Ni^{2+}$-NTA coated glass surface, and a probe bead was attached to $F_1$ via biotin-streptavidin interaction.

# Figure 2

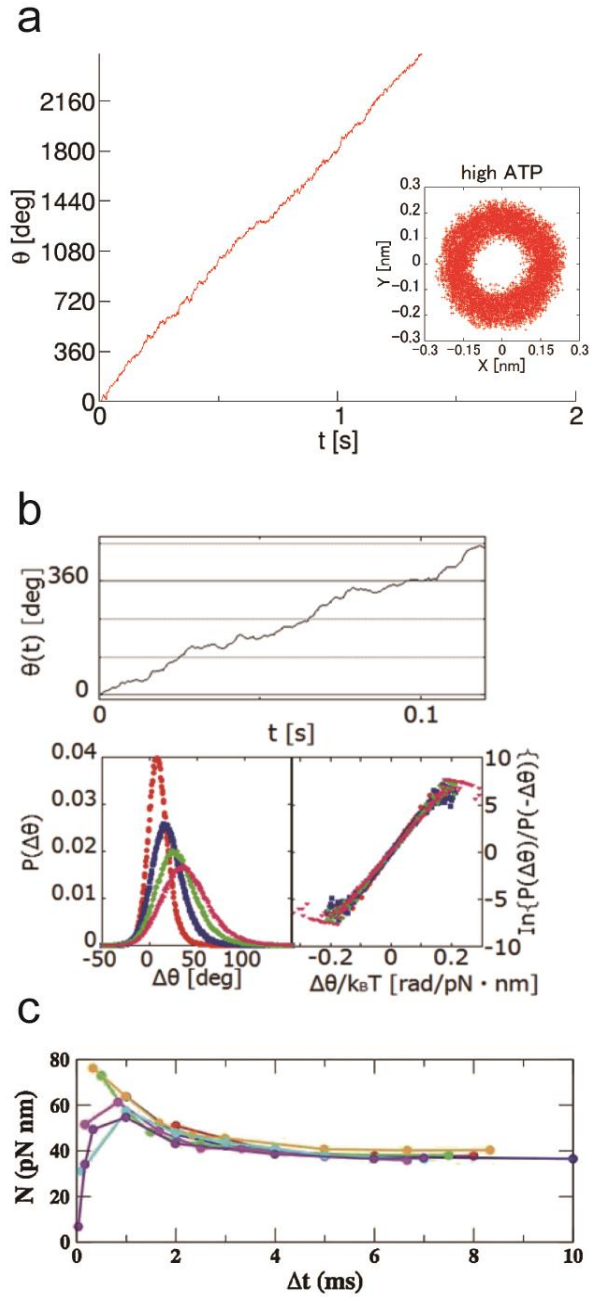

**Fig. 2** Fluctuation theorem (Eq. (1)) applied to $F_1$ (Hayashi et al. 2010). **a** An example of the time course of the rotary angle ($\theta$) in a high [ATP] condition (1 mM), calculated from the center positon of a rotated bead (inset). **B** Magnified figure of the time course ($\theta$) (top) in which the fluctuating behavior of the rotated bead was observed. The probability distribution $P(\Delta\theta)$ of $\Delta\theta(= \theta(t+\Delta t) - \theta(t))$ (bottom, left) and $\ln[P(\Delta\theta)/P(-\Delta\theta)]$ as a function of $\Delta\theta/k_BT$ for the cases $\Delta t$ = 2.5 ms (red), 5 ms (yellow), 7.5 ms (green) and 10 ms (blue). **b** $N(=k_BT \ln[P(\Delta\theta)/P(-\Delta\theta)]/\Delta\theta)$ as a function of $\Delta t$ for the recording rates, 500 fps (red),

1000 fps (blue), 2000 fps (green), 3000 fps (orange), 6000 fps (pink), 10000 fps (aqua), and 30000 fps (violet). After a certain amount of relaxation time, $N$ reaches a constant value.

# Figure 3

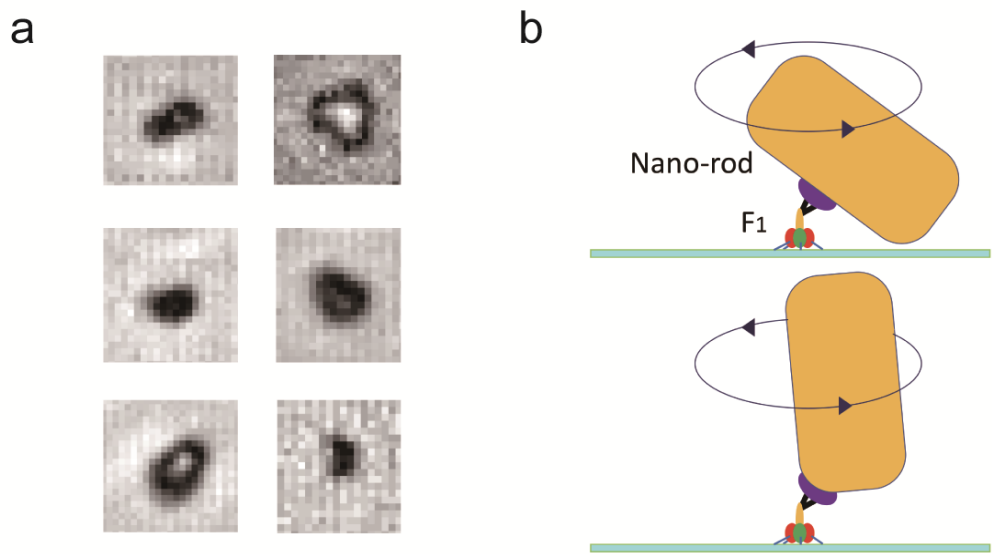

**Fig. 3** Difficulty in calculating the friction coefficient ($\Gamma$) of a rotated bead. **a** Example micrographs of probe beads. The shapes of beads were sometimes irregular. **b** Possible attachments of nano-rods to $F_1$. The attachment affects the value of $\Gamma$.

**Figure 4**

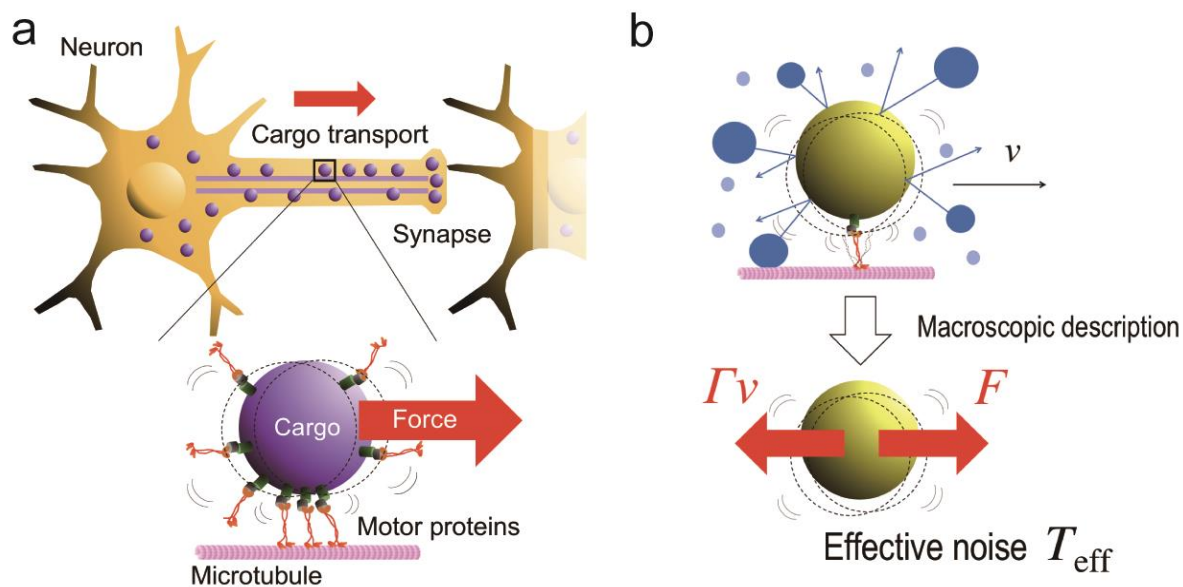

**Fig. 4** Cargo transport in neurons. **a** Schematics of cargo transport. Cargo is transported by motor proteins, such as kinesin and dynein, on microtubules in the axons of neurons. Cargo fluctuates due to thermal noise and intrinsic noise caused by collisions with other organelles and the cytoskeleton while in directional motion. Multiple motors are considered to transport a single cargo element cooperatively. **b** To represent the intrinsic noise caused by collision of the cargo with other vesicles and stochastic motion of the motors in addition to thermal noise in cells, the idea of effective temperature was introduced in Ref. (Hayashi et al. 2017).

# Figure 5

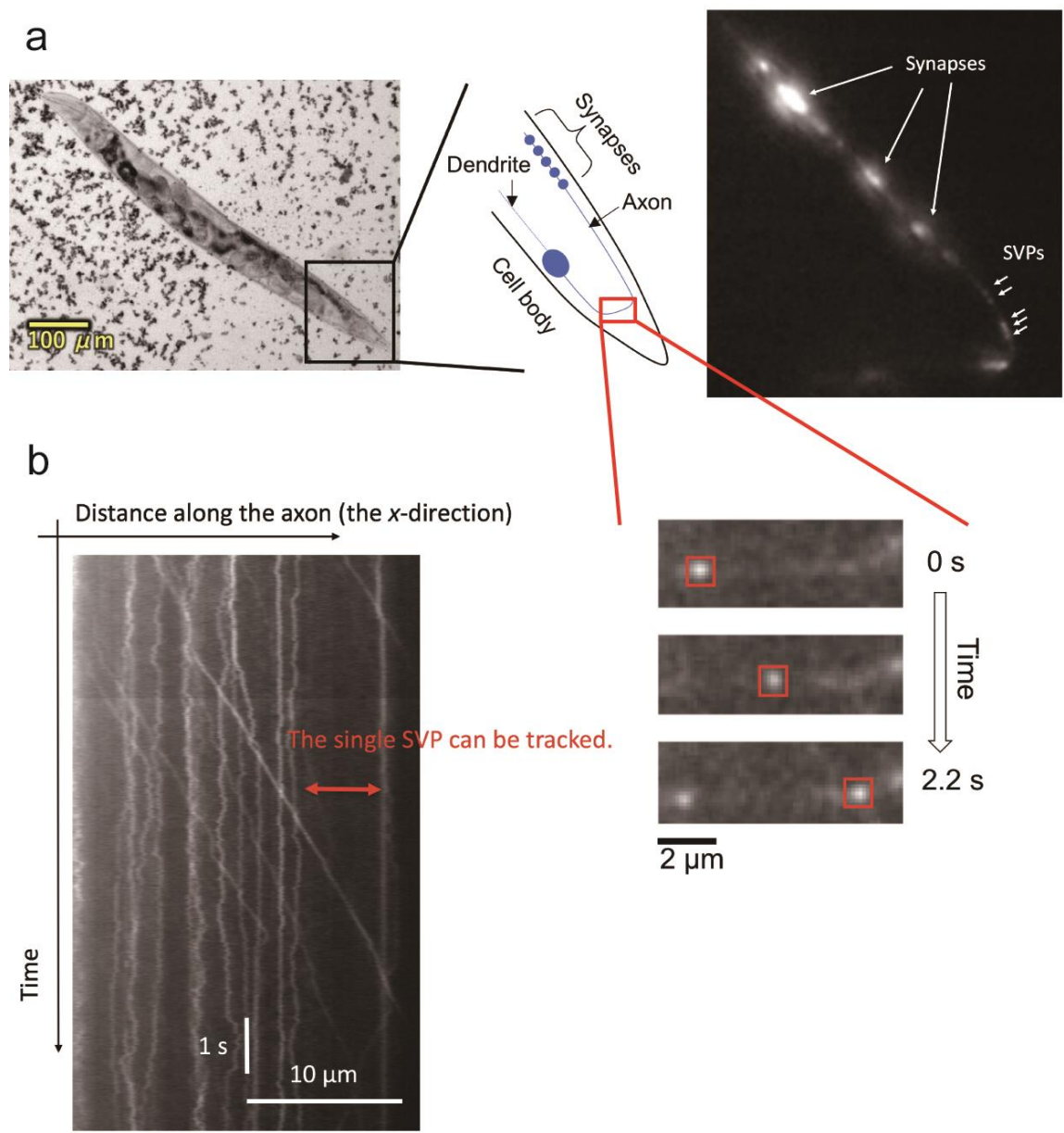

**Fig. 5** Fluorescence observation of synaptic vesicle precursors (SVPs) in DA9 neurons of *C. elegans* (Hayashi et al. 2018). **a** Bright-field micrograph of a *C. elegans* worm whose body was fixed in a chamber with high-viscosity medium and anesthesia (left). Magnified view around the anus of the worm (middle). GFP-labelled SVPs were tracked in the axon of the DA9 neuron (right). **b** Kymograph of SVPs. There were many SVPs in the axons. The kymograph represents the motion of SVPs along the direction of the axon as a function of time. Each bright line represents the motion of an SVP. Several SVPs moved anterogradely. There were time intervals during which a single SVP could be tracked (red arrows), even though the axon was crowded with many SVPs. We investigated such rare time intervals.

# Figure 6

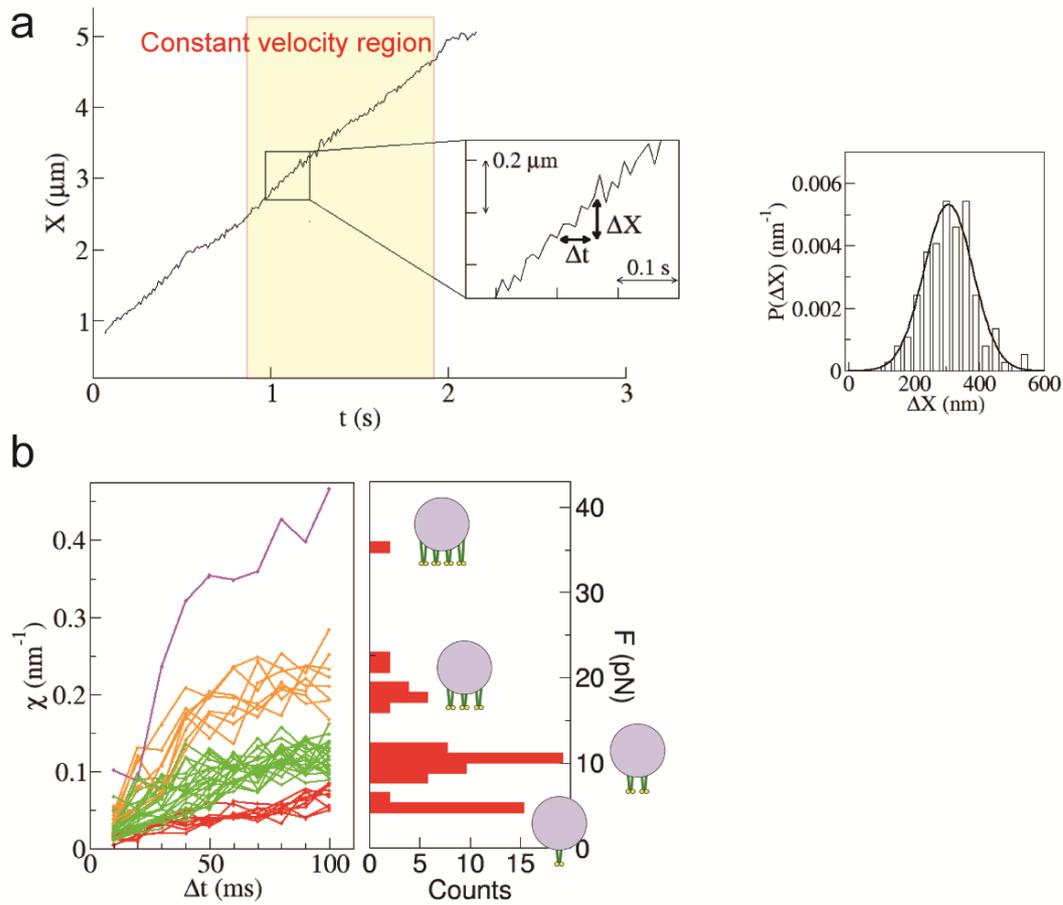

**Fig. 6** Modified fluctuation theorem (Eq. (5)) applied to SVP transport in the axon of *C. elegans* (Hayashi et al. 2018). **a** Center position (*X*) of an SVP in anterograde transport as a function of time (left). Our analysis target was a constant-velocity interval that lasted longer than 0.5 s. When the recording speed was high enough (100 frames per second), fluctuation of *X* was observed (inset). Here, $\Delta X = X(t + \Delta t) - X(t)$. The probability distribution ($P(\Delta X)$) of $\Delta X$ for the case of 100 ms (right). The distribution was well fitted by a Gaussian function (thick curve). **b** $\chi$ (eqn. (6)) as a function of $\Delta t$ for SVP anterograde runs in wild-type *C. elegans* (n = 40 SVPs from 33 different worms). There are four clusters as determined by cluster analysis (see Method in Ref. (Hayashi et al. 2018)). Each color denotes a cluster, representing different force-producing units (FPUs). See Method of Ref. (Hayashi et al. 2018) for a conversion from $\chi$ to *F* based on Eq. (7).

**Figure 7**

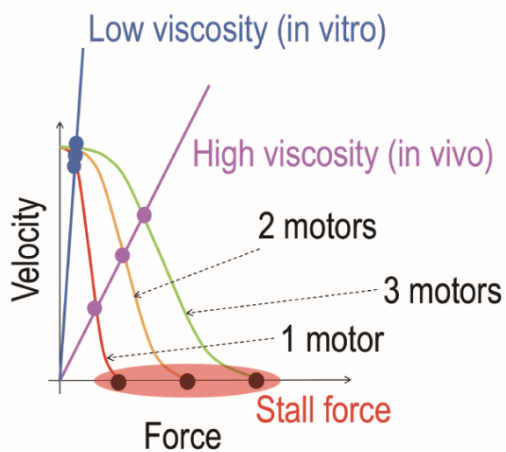

**Fig. 7** Force-velocity relation of motors. Originally, the force-velocity relation was introduced through *in vitro* single-molecule experiments on kinesin (Schnitzer et al. 2000). A similar relation for *in vivo* cargo transport was assumed in Ref. (Hayashi et al. 2017). Drag force ($\Gamma v=F$) values of 1 motor (red), 2 motors (yellow), and 3 motors (green) are considered discrete in the high viscosity case, similarly to the values of stall force.

**Figure 8**

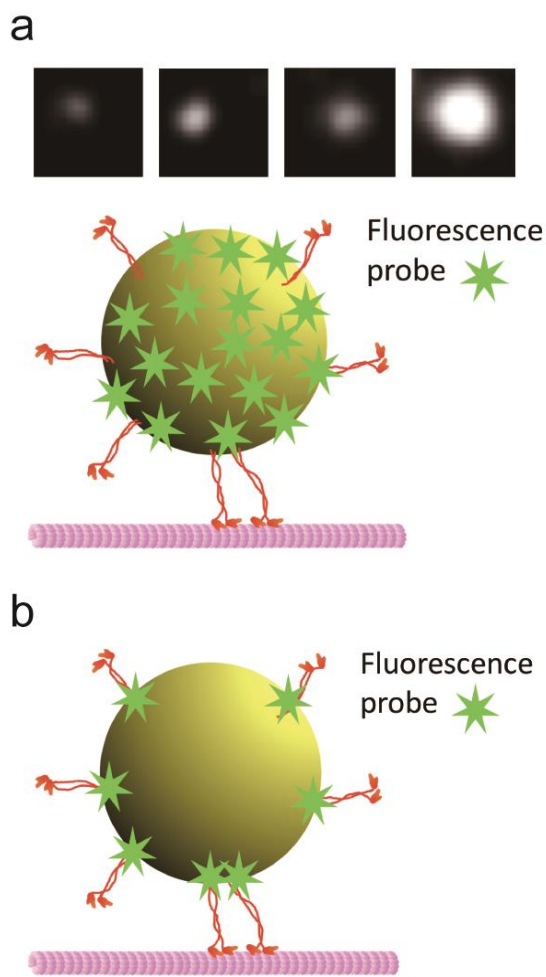

**Fig. 8** Observable quantities such as fluorescence intensity other than $\chi$ (eqn. (6)). **a** Fluorescence intensities of cargo (Hayashi et al. 2017) were different for different cargo sizes when fluorescence probes were uniformly attached to cargo surfaces. **b** Number of active motors (e.g., two for the schematics) that haul cargo cannot be counted from fluorescence intensities, even when fluorescence probes are directly attached to motors, because motors that do not haul a cargo also emit fluorescence.

# Figure 9

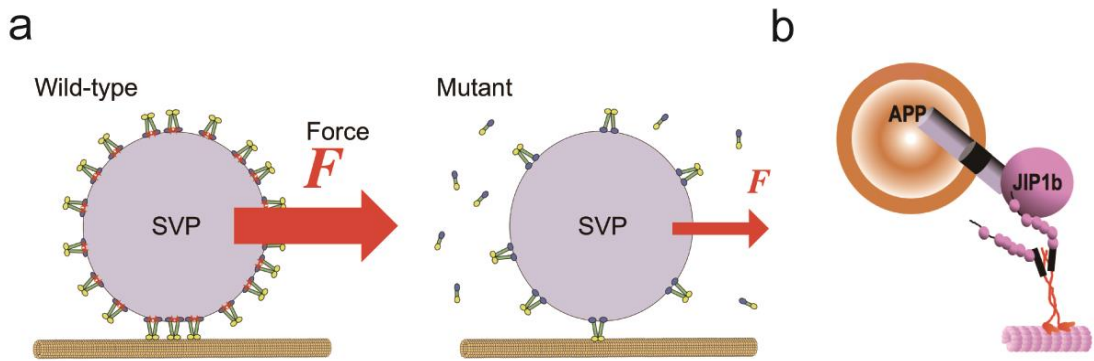

**Fig. 9** Motor number is an important indicator of axonal cargo transport. **a** The number of motors was found to decrease in an *arl-8*-deleted mutant (Hayashi et al. 2018), in which decrease in synapse density and mislocation of synapses occurred (Niwa et al. 2016). **b** A decrease in velocity was observed when the adaptor protein jip1b was deleted (Chiba et al. 2014). The physical reason for this decrease warrants future investigation.